\documentclass{aa}
\usepackage{natbib}
\bibpunct{(}{)}{;}{a}{}{,}
\usepackage{graphicx}
\usepackage{txfonts}
\usepackage[colorlinks,
            linkcolor=red,
            anchorcolor=blue,
            citecolor=blue]{hyperref}

\begin{document}

   \title{Hot subdwarf B stars with neutron star components II: Binary population synthesis}


   \author{You Wu
          \inst{1,2,3},
          Xuefei Chen\inst{1,2,4},
          Hailiang Chen\inst{1,2,4},
          Zhenwei Li\inst{1,2,3}
          \and
          Zhanwen Han\inst{1,2,4}
          }
   \authorrunning{Y. Wu
          \inst{1,2,3},
          X. Chen\inst{1,2,4},
          H. Chen\inst{1,2,4},
          Z. Li\inst{1,2,3}
          \and
          Z. Han\inst{1,2,4}
          }
   \institute{
   $^1$Yunnan Observatories, Chinese Academy of Sciences,
            Kunming 650216, China; youwu@ynao.ac.cn, cxf@ynao.ac.cn\\
           $^2$ Key Laboratory for the Structure and Evolution of Celestial Objects, Chinese Academy of Sciences, Kunming 650216, China\\
            $^3$ University of the Chinese Academy of Science, Beijing 100049, China\\
            $^4$ Center for Astronomical Mega-Science, Chinese Academy of Sciences, 20A Datun Road, Chaoyang District, Beijing, 100012, China\\
             }

   \date{}


  \abstract
   {Subdwarf B stars (sdBs) play a crucial role in stellar evolution, asteroseismology,
   and far-UV radiation of early-type galaxies,
   and have been intensively studied with observation and theory.
  It has theoretically been predicted that sdBs with neutron star (NS) companions exist in the Galaxy,
   but none have been discovered yet. This remains a puzzle in this field. In a previous study (hereafter Paper I), we have studied the formation channels of sdB+NS binaries
from main-sequence (MS) stars plus NS binaries by establishing a model grid, but it is still unclear how these binaries consisting of MS stars and NS binaries came to be in the first place.}
   {We systematically study the formation of sdB+NS binaries from their original zero-age main-sequence progenitors. We bridge the gap left by our previous study in this way. We obtain the statistical population properties of sdB+NS binaries and provide some guidance for observational efforts.}
{
We first used Hurley's rapid binary evolution code BSE to evolve $10^7$ primordial binaries to the point where the companions of NS+MS, NS+Hertzsprung gap (HG) star, and NS+Giant Branch (GB) star binaries have just filled their Roche lobes.
Next, we injected these binaries into the model grid we developed in Paper I to obtain the properties of the sdB+NS populations.
We adopted two prescriptions of NS natal kicks: the classical Maxwellian distribution with a dispersion of $\mathrm{\sigma = 265 kms^{-1}}$,
and a linear formula that assumes that the kick velocity is associated with the ratio of ejected to  remnant mass.
Different values of $\alpha_{\rm CE}$,
where $\alpha_{\rm CE}$ is the common-envelope ejection efficiency,
were chosen to examine the effect of common-envelope evolution on the results. }
{In the Galaxy, the birthrate of sdB+NS binaries is about  $10^{-4}\mathrm{yr^{-1}}$
and there are $\sim 7000-21000$ such binaries. This
contributes 0.3-0.5\% of all sdB binaries in the most favorable case.
Most Galactic sdB+NS binaries ($\gtrsim 60\%$) arise from the channel of stable mass transfer.
The value of $\alpha_{\rm CE}$ has little effect on the results,
but when we use the linear formula prescription of NS natal kick, the number and birthrate doubles in comparison to the results we obtained with the Maxwellian distribution.
The orbital periods of sdB+NS binaries from different formation channels differ significantly, as expected. This results in two peaks in the radial velocity (RV) semi-amplitude distribution:
$100-150 {\rm {kms^{-1}}}$ for stable mass transfer,
and $400-600 {\rm {kms^{-1}}}$ for common-envelope ejection.
However, the two sdB+NS binary populations exhibit similar delay-time distributions,
which both peak at about 0.2Gyr. This indicates that Galactic sdB+NS binaries are born in very young populations, probably in the Galactic disk.
The sdB+NS binaries produced from the common-envelope ejection channel
are potential  sources of strong gravitational wave radiation (GWR),
and about $\sim 100-300$ could be detected by the Laser Interferometer Space Antenna (LISA)
with a signal-to-noise ratio of 1.}
{Most sdB+NS binaries are located in the Galactic disk with small RV semi-amplitudes.
SdB+NS binaries with large RV semi-amplitudes are expected to be strong GWR sources, some of which could be detected by LISA in the future.}

               
\titlerunning{SdB with NS components II: Binary population synthesis}         
   \maketitle
%

\section{Introduction}
Hot subdwarf B (sdB) stars, also known as extreme horizontal branch stars,
are He-core burning stars
with very thin hydrogen-rich envelopes ($\lesssim 0.02 \,M_\odot$) \citep{Heber1986}.
These objects exhibit high effective temperatures ($\mathit{T}_{\mathrm{eff}}\approx 20000- 40000\mathrm{K}$) and high surface gravities ($\log \mathit{g} {\rm (cm s^{-2})}$ 
$\sim$ 5.0 - 6.5) \citep{Heber2016}.
SdBs play a crucial role in stellar evolution and asteroseismology and heavily influence the far-UV radiation of early-type galaxies \citep{1996ApJ...471L.103C,Charpinet,1997MNRAS.285..640K,2000ApJ...532..308B,2007MNRAS.380.1098H}. They have consequently attracted intensive interest from observers and theorists alike \citep[e.g.,][]{Han2002,Han2003,2010A&A...519A..25G,Chen2013,2018MNRAS.473..693V}.
A large portion of sdBs are binaries \citep{Maxted,Napiwotzki2004}.
\citet{Han2002,Han2003} developed a binary model for the formation of sdBs,
in which sdB binaries originate from either stable Roche-lobe overflow (the RLOF channel)
or from common-envelope ejection (the CE channel),
while single sdBs arise from mergers of two He white dwarfs (WDs) \citep[see also][]{Webbink1984}.
The model nicely reproduced properties of short-period sdB+WD binaries and single sdBs.
Following the discovery of sdB binaries with long orbital periods \citep{2012MNRAS.421.2798D},
\citet{Chen2013} revisited the period distribution by considering the mass-luminosity relation of the cores of giant stars.
This period distribution has been confirmed by recent studies \citep{2019MNRAS.482.4592V}.

We here study sdBs with neutron star (NS) or black  hole (BH) companions. Theoretically, these objects can be produced through binary evolution, and they may eventually evolve to become WD+NS or WD+BH binaries.
For example, sdB+NS binaries can be produced from the evolution of systems that eventually become HeCO-WD+NS binaries\citep{Podsiadlowski_2002}.
\citet{Tauris2011} also clearly demonstrated the existence of an intermediate evolutionary phase
of sdB+NS binaries when they modeled the formation of the massive millisecond-pulsar binary PSR J1614-2230.
\citet{Nelemans2013} performed binary population synthesis (BPS) studies for sdBs
and predicted that about 1\% of sdBs in the Galaxy have NS companions and 0.1\% have BH companions.
They considered the effect of asymmetric kicks that are imparted on NSs by their host supernovae,
but ignored the details of the formation process of sdB+NS binaries.

\citet{Geier2011} proposed the project Massive Unseen Companions to Hot Faint Under-luminous Stars from SDSS (MUCHFUSS), with the aim of finding massive compact companions, such as massive WDs (> $1.0 \,M_\odot$),
NSs, or stellar mass BHs, of hot sdB/O stars.
So far, they have detected 53 candidates in total,
but did not find any evidence for the existence of NS or BH companions to sdB/O stars \citep{Geier2015,Geier2017a}. They proceeded to suggest 1.5\% as an upper limit for the fraction of close sdB+NS or sdB+BH binaries.

In order to better understand the observations of MUCHFUSS,
 \citet{Wuyou} (hereinafter Paper I) systematically investigated the formation of sdB+NS binaries
 by establishing a model grid for a series of main-sequence star (MS) + NS binaries.
They showed that sdB+NS binaries may be produced either from the RLOF channel or from the CE channel.
For those that evolved through the RLOF channel, the orbital period ranges from several days to more than 1000 days. The highest radial velocity (RV) semi-amplitude $K$ is about $150\mathrm{kms^{-1}}$.
For those from the CE channel, the orbital period is very short, with high values of $K$ of up to $800\mathrm{kms^{-1}}$. Gravitational wave radiation (GWR) may cause them to merge on a timescale of only $ {\rm \text{some megayears}}$.
This implies that sdB+NS binaries from both channels are difficult to discover.

The main goal of this study is to obtain the properties of Galactic sdB+NS binaries,
such as the number, period distribution, dependence on the age, and 
the uncertainties of binary evolution and NS natal kicks,
by combining a new BPS study with the model grid established in Paper I.
The remaining part of the paper is structured as follows.
In Sect. 2 we present the formation channels of sdB+NS binaries from primordial binaries.
Sect. 3 contains the methods we used in our BPS simulation and the parameter settings of our binary evolution.
The simulation results are shown in Sect. 4, and the main conclusions are summarized in Sect. 5.

\section{Formation scenario for sdB+NS binaries}
Figure 1 illustrates the evolution of a primordial binary into an sdB+NS binary.
The primary (the initially more massive component) evolves faster and fills its Roche lobe first, then transfers material to the secondary.
The mass transfer can be dynamically stable or unstable, depending on the initial mass ratio.
For binaries with stable mass transfer, the companion star (i.e., the initially less massive component) is relatively massive in comparison to its counterparts in systems with dynamically unstable mass transfer because of the constraint of the mass ratio for stable mass transfer. They subsequently accrete some of the material during the mass-transfer phase, which increases their mass. Subsequently, the produced NS+MS binaries have relatively large MS companions, that is, the MS mass $ \gtrsim 10M_{\odot}$. Because of the high mass ratio of these binaries, they ultimately undergo unstable mass transfer and enter the CE phase when the secondary fills its Roche lobe.
If the binaries survive this CE phase, the secondaries evolve
into massive He stars with very high effective temperatures. These are not sdB stars.

For the case of dynamically unstable mass transfer, the binary enters a CE phase and evolves into a He+MS binary after the ejection of its CE.
The He star further evolves and leaves behind an NS after supernova explosion if its mass is in the appropriate range.
The secondary (the present MS star) evolves and subsequently fills its Roche lobe.
What happens next depends upon the mass of the secondary
and the orbital period.
The mass transfer may be dynamically stable or unstable (to form a CE).
Both cases may produce sdB+NS binaries, according to the results of Paper I (and this is referred to here as the CE+CE channel and CE+RLOF channel, respectively). In our calculations, we assume that the system forms an sdB if helium can be ignited successfully, steady burning in the center of stars is maintained after the mass-transfer process, the envelope is stripped to such an extent that its mass is relatively low, and if the effective temperature is lower than $4\times10^{4}$\;K.

By investigating the evolution of a series of NS+MS binaries with the MESA code \citep{Paxton2011,Paxton,Paxtona}, we obtained the parameter space for producing sdB+NS binaries through the RLOF and CE channels in Paper I, as shown in the companion mass - orbit period ($M_{2}$-$\lg P_{\rm m}$) diagram of Fig.2, where $M_{\rm 2}$ is the secondary mass and $P_{\rm m}$ is the orbital period when the secondary fills its Roche lobe.
In the calculation, we followed the detailed evolution of the mass-transfer rate for each binary. In systems where the mass-transfer rate increases dramatically in a short time and becomes higher than a critical rate, that is, $10^{-4}M_{\odot}yr^{-1}$,
we assume that the binary system undergoes dynamically unstable mass transfer. 

In the CE channel, we adopt
the standard energy prescription for CE evolution (see Sect. 3.3 for details).
The binding energy here considers the combined effects of gravitational energy and internal energy and is slightly different from that of Paper I, where these two sources of energy were considered separately (see their Eq. (1)). In Paper I, we used a different approach for
calculating the CE evolution, therefore this leads to an inconsistency in the treatment of the two CE phases in the present study. To resolve this problem, we repeated the CE evolution in Paper I, adopting the same prescriptions for the CE that are used in this study, including the calculation of $\lambda$ and different $\alpha_{\rm CE}$ values, and substitute the results of Paper I with those of this new set of calculations.

The CE channel and the stable RLOF channel in Fig.2 are separated by a gap.
When the orbital periods of the NS+MS, NS+HG, and NS+GB binaries lie in this gap,
the mass-transfer process is dynamically unstable and a CE forms,
but the CE cannot be ejected successfully because its binding energy is high.
The parameter space lies in a fairly narrow range when the mass of the companion star is lower than $2.5 \,M_\odot$: He can only be ignited subsequently when the donors start mass transfer close to the tip of the red giant branch
if the donors have degenerate He cores.
There is a transition from degenerate to non-degenerate He cores
when the mass of the companion stars lies in the range of 2.0 to $2.5 \,M_\odot$,
resulting in a small parameter space for producing sdBs in this mass range.

\begin{figure}
   \centering
   \includegraphics[width=0.4\textwidth]{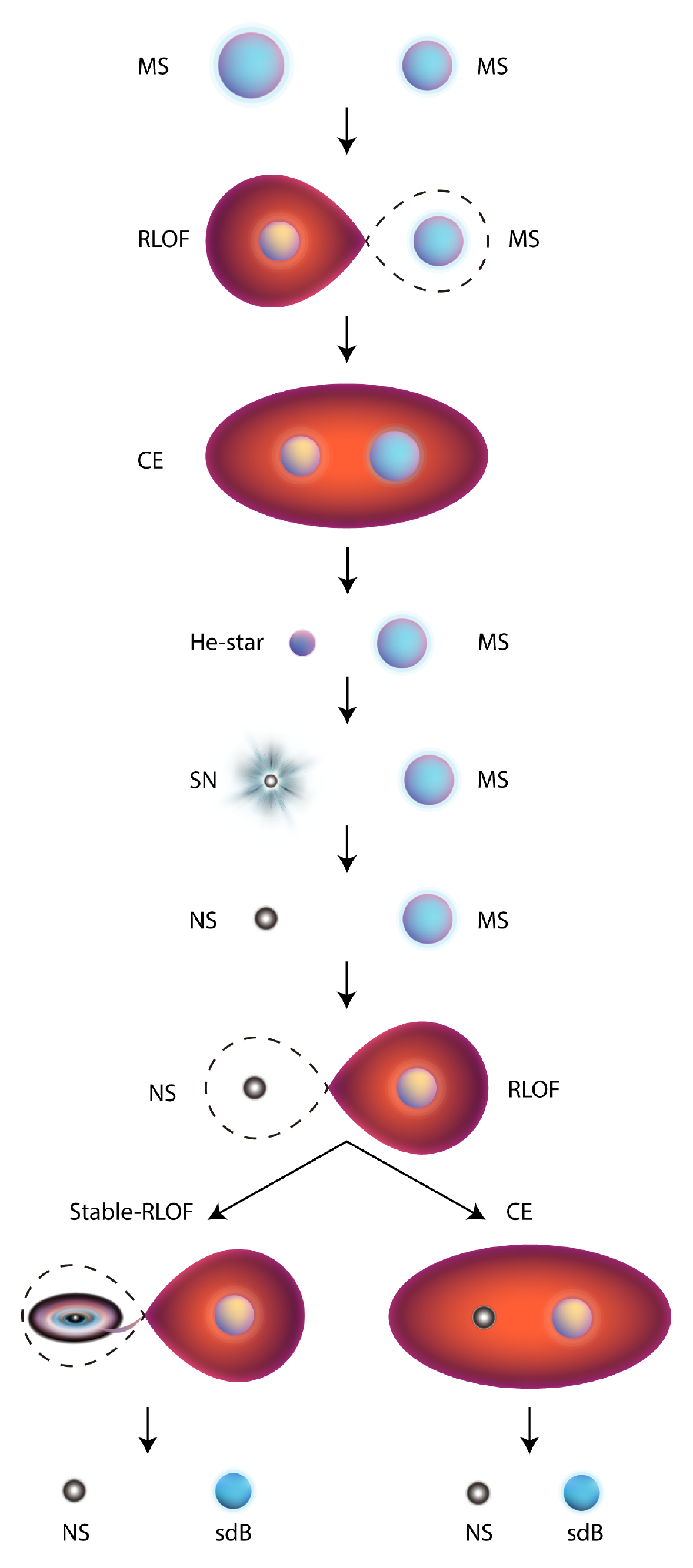}
      \caption{Illustration of the formation of sdB+NS binaries.
      For the first mass-transfer process, we do not show the stable-RLOF case in the figure because the NS+MS binaries produced by stable RLOF do not contribute to the production of sdB+NS binaries (see Sect. 2 for details). }
         \label{11111}
\end{figure}

 \begin{figure}
   \centering
   \includegraphics[width=\hsize]{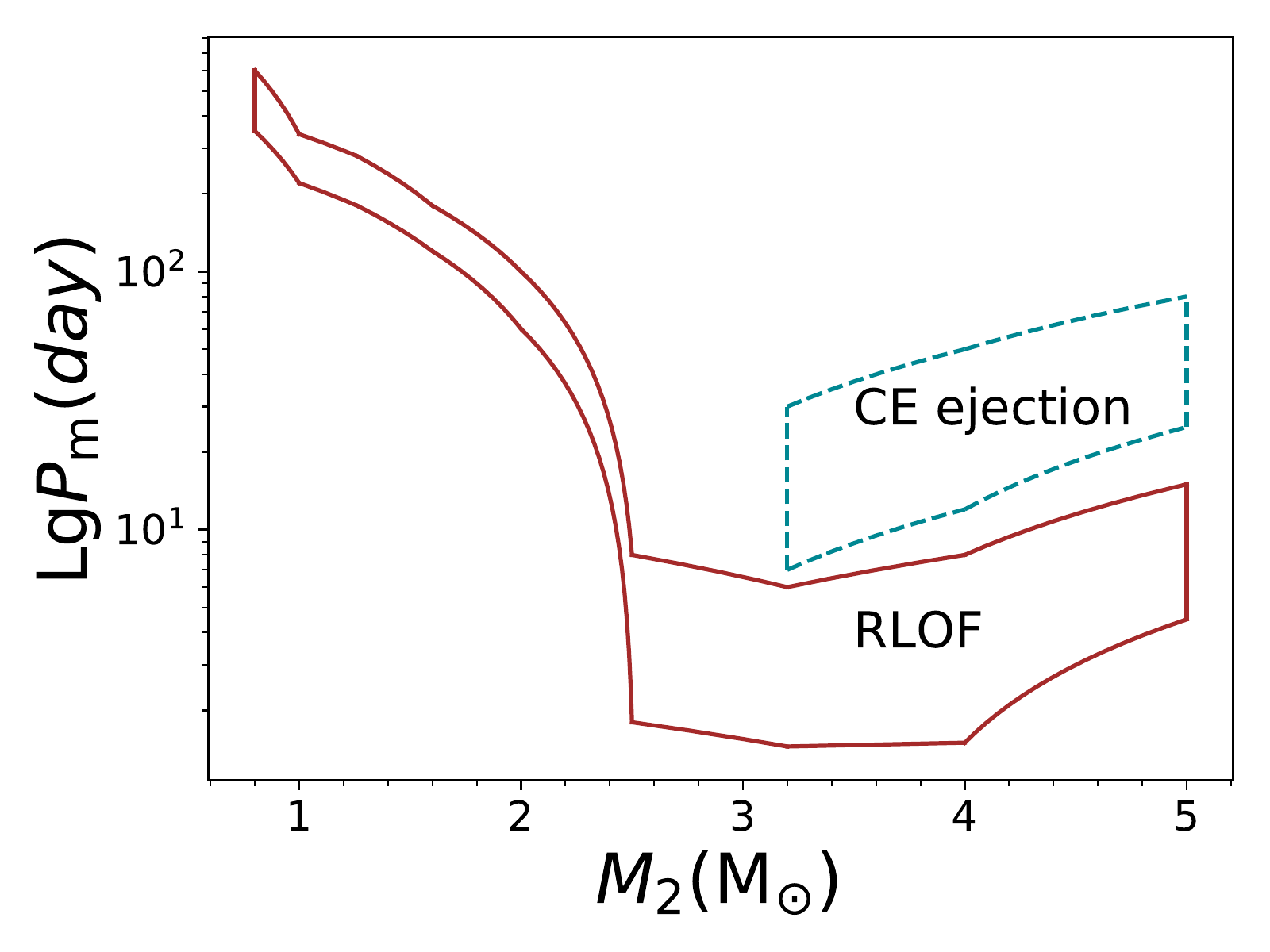}
      \caption{Parameter spaces for NS+MS, NS+HG, and NS+GB binaries that can produce sdB+NS binaries, where $M_2$ is the companion mass and $P_{\rm m}$ is the orbital period when the companion fills its Roche lobe.
       $M_2$ ranges from $0.8 M_{\odot}$ to $5.0 M_{\odot}$ (MS stars with mass $>5.0M_{\odot}$ could soon evolve to high-mass X-ray binaries through accretion by the stellar wind).
      The regions enclosed with the solid and dashed lines are for the stable RLOF channel
      and the CE ejection channel, respectively. In the CE channel, $\alpha_{\rm CE} = 1$.  The results for $\alpha_{\rm CE} = 0.5, 2,$ and 3 are shown in Fig.3. }
         \label{space}
   \end{figure}

\section{Simulation of the binary population synthesis}
In order to obtain the properties of Galactic sdB+NS binaries,
such as the number, period distribution, dependence on the age or metallicity,
we performed eight sets of BPS simulations.
For each simulation set, we generated $10^7$ primordial binaries by a Monte Carlo method,
and then we used the rapid binary stellar evolution code BSE \citep{2000Hurley,2002Hurley} to evolve these binaries to obtain a sample of binaries consisting of an NS and a non-degenerate companion (i.e., MS, HG, and GB),
in which the companion stars just fill their Roche lobes and start transferring mass to the NSs.
We use $M_{\rm 2}$ to denote the companion mass hereinafter for convenience.
We then combined these binaries with the model grid and obtained the information for the sdB+NS binaries.

We study Population I stars with a metallicity of $Z=0.02$.
The physics inputs for the eight sets are summarized in Table 1, and the details are described below.

\begin{table}
\caption{\label{t7} Parameters for the BPS simulation in our study.}
\centering
\begin{tabular}{ccc}
 \hline \hline
$\mathrm{Set}$ & $\mathrm{Natal\, kick}$  & $\alpha_{\mathrm{CE}}$ \\
\hline
 1   & $Vk_{\rm{linear}}$ & 0.5   \\
 2   & $Vk_{\rm{linear}}$   & 1           \\
 3   & $Vk_{\rm{linear}}$    & 2    \\
 4   & $Vk_{\rm{linear}}$    & 3    \\
 5   & $Vk_{\rm{\sigma = 265 km s^{-1}}}$ &0.5  \\
 6   & $Vk_{\rm{\sigma = 265 km s^{-1}}}$ &1\\
 7   & $Vk_{\rm{\sigma = 265 km s^{-1}}}$ &2\\
 8   & $Vk_{\rm{\sigma = 265 km s^{-1}}}$ &3\\
  \hline  \hline
\end{tabular}
\tablefoot{$Vk_{\rm{linear}}$ and $Vk_{\rm{\sigma = 265 km s^{-1}}}$ are for the prescriptions of NS natal kicks from CCSNe, i.e., Eqs. (4) and (3), respectively.  $\alpha_{\mathrm{CE}}$ is the CE ejection efficiency.}
\end{table}

\subsection{Initial distribution for binary parameters}
To generate the primordial binaries,
we need the initial distributions for the binary parameters as inputs in the Monte Carlo simulation,
that is, primary mass, mass ratio, and orbital separation.
We describe these as follows.

(i) We adopted the initial mass function of \citet{1979Miller} for the distribution of the primary mass $M_{1}$, as given by \citet{1989ApJ...347..998E},
\begin{equation}
   M_{1}= \frac{0.19X}{(1-X)^{0.75}+0.032(1-X)^{1/4}}
   ,\end{equation}
where $X$ is a random number in the range from 0 to 1.

(ii) The initial mass-ratio $q$ (The ratio of the primary to the secondary) distribution is taken as a constant distribution, that is, $n(1/q)=1$.

(iii) The distribution of the initial separation is a uniform distribution in log $a$ for wide
binaries and a power-law distribution at close separation \citep{1998MNRAS.296.1019H},
\begin{equation}
   an(a)=\left\{\begin{matrix}
\alpha _{\rm{sep}}(\frac{a}{a_{0}})^{m},\quad a\leqslant a_{0}; \\
\alpha _{\rm{sep}},\quad a_{0}< a< a_{1},
\end{matrix}\right.
   \end{equation}
where $a$ is orbital separation, $\alpha _{\rm{sep}}\approx 0.07,\;  a_{0}=10\mathrm R_{\odot},\;  a_{1}=5.75\times 10^{6}\mathrm R_{\odot}\; \mathrm{, and}\; m\approx 1.2$.
This distribution yields approximately 50\% of the binaries with orbital periods shorter than 100 yr.

\subsection{NS formation and their kicks}
It is known that NSs may be produced from core-collapse supernovae (CCSNe),
electron-capture supernovae (ECSNe) \citep{Miyaji1980}, or accretion-induced collapse (AIC)
of an oxygen-neon-magnesium WD (ONeMg WD) \citep{Nomoto1984,Nomoto1987,1987Natur.329..310M}.
In our study, the CCSNe occurs for a star with a He core mass $M_{\rm He}>2.25M_{\odot}$, leaving an NS or a black hole after supernovae.
When the remnant mass (for the calculation of the remnant mass, we refer to \citealt{2000Hurley}.) was lower than $3.0M_{\odot}$, we assumed that an NS was left (according to the Tolman-Oppenheimer-Volkoff limit in \citealt{1939PhRv...55..374O}).
Similar to \citet{2018Chruslinska}, we assumed that a star with a He-core mass in the range of $1.83M_{\odot} < M_{\mathrm{He}}<2.25M_{\odot}$  produces an NS from ECSNe when the remnant mass approaches the Chandrasekhar mass limit.
In the AIC formation scenario, the mass accumulation efficiency of the WDs was computed as follows.
When the mass-transfer rate exceeded a critical mass-transfer rate, we employed the optically thick wind model proposed
by \citet{1996ApJ...470L..97H}.
In the stable burning region of H or He, the mass-accumulation efficiency was assumed to be 1.
When the mass-transfer rate was lower than the minimum accretion rate of stable burning, we separately computed the mass accumulation efficiency with
 the prescriptions of \citet{1999Hachisu} and \citet{2004Kato} for H and He burning. These prescriptions have been widely
 used in the population synthesis study of type Ia supernovae \citep[e.g.,][]{2004MNRAS.350.1301H,2009MNRAS.395..847W,2018Liu}.

We here simply assumed an NS birth mass of $1.4 M_{\odot}$ for convenience.
However, we recall that the NS mass from different formation scenarios may be different.
The mass of NSs produced by ECSNe and AIC is suggested to range from $1.25 M_{\odot}$ to $1.27 M_{\odot  }$ \citep{1996ApJ...457..834T,2010ApJ...719..722S,2011BASI...39....1V}. \citet{2006ApJ...644.1063D} demonstrated that the AIC can form NSs with a mass of $\sim 1.4 M_{\odot}$.
\citet{2010PhRvL.104y1101H} and \citet{2010A&A...517A..80F} proposed $1.366 M_{\odot}$ and $1.347 M_{\odot}$ as the final baryon mass of NSs from the $8.8 M_{\odot}$ ECSNe progenitor, respectively.
Based on the distribution of NS masses in observations, most NS masses are concentrated near $\sim 1.4 M_{\odot}$\citep{2012ARNPS..62..485L}.

The NS natal kick is crucial for the formation of binaries with NS companions because most binaries may be destroyed by SNe when NSs are born.
Even when binaries survive the SNe,
their orbital periods and eccentricities are still highly sensitive to NS natal kicks.
It is commonly considered that NSs from ECSNe or AIC receive much weaker natal kicks than those from CCSNe \citep{Pfahl2002,2004ApJ...612.1044P,2004PhRvL..92a1103S,2006ApJ...644.1063D},
but the uncertainty on this question is large.
In our simulation, we assumed that no natal kicks were imparted onto NSs produced by ECSNe or AIC,
and we adopted two prescriptions for the natal kick velocity for NSs from CCSNe.

(i) \citet{2005Hobbs} suggested a well-fit Maxwellian velocity distribution with a dispersion of $\mathrm{\sigma = 265 kms^{-1}}$ based on an analysis of proper motion measurements of a variety of pulsars, 
\begin{equation}
P(v_{\mathrm k})=\sqrt{\frac{2}{\pi }}\frac{v_{\mathrm k}^{2}}{\sigma^{3}}\mathrm e^{-{v_{\mathrm k}^{2}} /2\sigma^{2}}.
\end{equation}

(ii) Inspired by SN explosion asymmetries, \citet{2016Bray,2018Bray} proposed a direct connection between the kick velocities and the ratio of the ejected mass of an SN explosion to NS mass, that is,

\begin{equation}
v_{\rm{k}}=\alpha(\frac{M_{\rm{eject}}}{M_{\rm{remnant}}})+\beta,
 \end{equation}
where $M_{\rm{eject}}$ is ejecta mass from the SN, $M_{\rm{remnant}}$ is the NS mass, and $\alpha$ and  $\beta$ are constants. \citet{2018Bray} suggested a best-fit kick with $\alpha=100$ and $\beta=-170$.
For convenience, we denote them with $Vk_{\rm{\sigma = 265 km s^{-1}}}$ and $Vk_{\rm{linear}}$.

 \begin{figure*}
   \centering
   \includegraphics[width=\hsize]{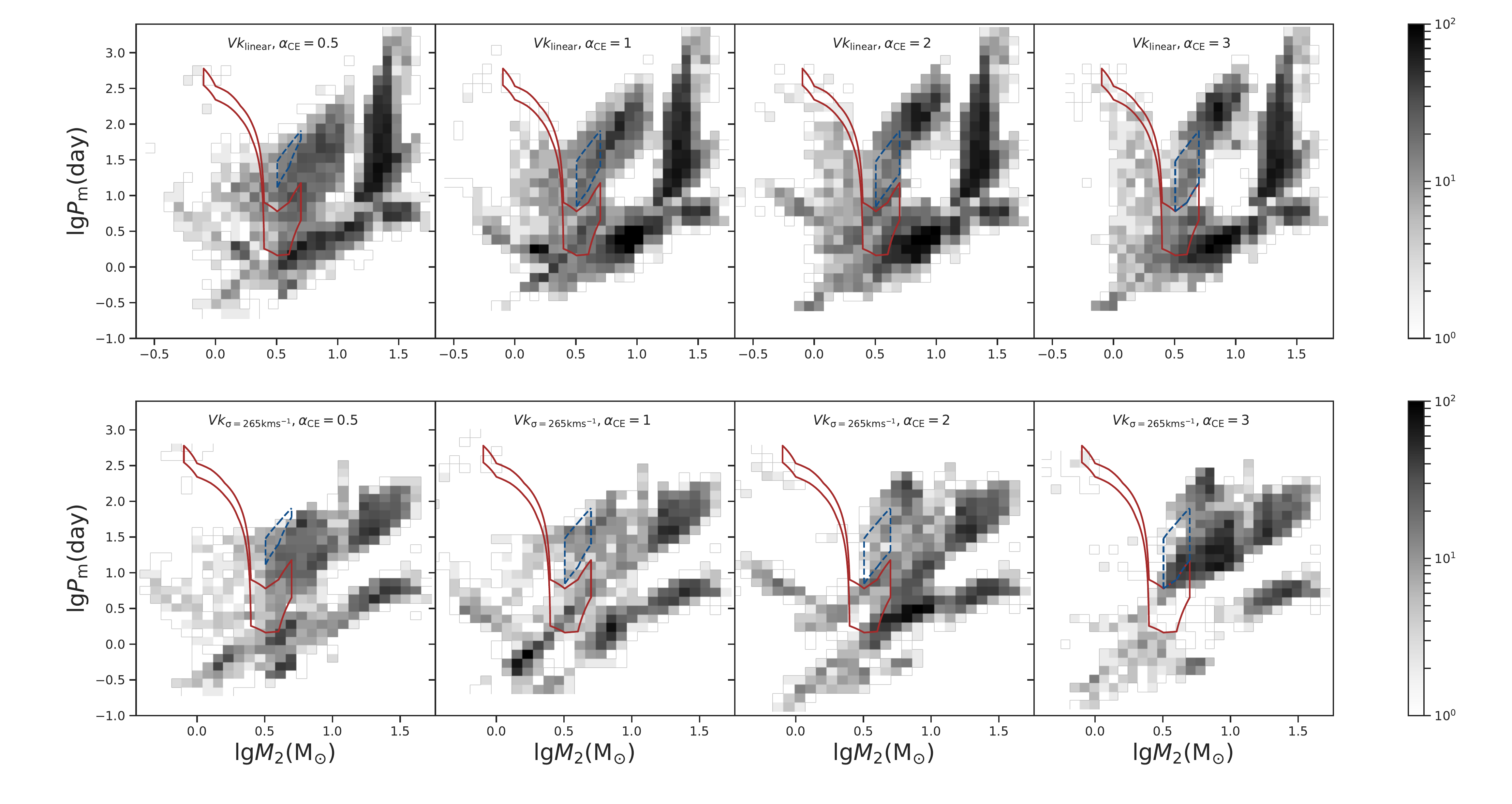}
      \caption{Number density distribution of NS+MS, NS+HG, and NS+GB binaries on the mass (${\rm log} M_{\rm 2}$) - period (${\rm log} P_{\rm m}$) plane
      when the MS, HG, and GB companion just fills its Roche lobe.
      The regions enclosed within the solid and dashed lines
      are parameter spaces for producing sdB+NS binaries from the RLOF channel
      and from the CE ejection channel, respectively.
      The prescriptions for NS kicks and the value of $\alpha_{\rm CE}$ are indicated in each panel.}
         \label{nctt}
   \end{figure*}

\subsection{Dynamical instability and CE evolution}
It is an unresolved problem for the binary interaction
whether the mass transfer is dynamically stable.
To determine the mass transfer stability,
we assumed that if the mass ratio at the onset of mass transfer is higher than a critical mass ratio $q_{\rm{c}}$ (mass ratio of the donor to the accretor),
the mass transfer is dynamically unstable and a CE forms soon after.
We set $q_{\rm c}$ to be 3.0 for donors on the MS, He MS, and in the He-core burning phase,
and 4.0 for donors during the HG.
When the donors were on the GB,
 the model of condensed polytrope was adopted \citep{1987Hjellming}, that is,
\begin{equation}
q_{\rm c}= 0.362+[3(1-M_{\rm c}/M_{\rm d})]^{-1},
 \end{equation}
where $M_{\rm d}$ and $M_{\rm c}$ are the mass and core mass of the donor, respectively.
In particular, $q_{\rm{c}}$ was set to be 0.784 for He stars in GB \citep[see][]{2002Hurley}. Note that the $q_{\rm{c}}$ given here is valid for the evolution before the formation of NS+MS binaries. The following evolution of NS+MS binaries depends on the detailed binary evolution calculation as described in section 2 and in Paper I.

The CE evolution is another unresolved problem in binary evolution (see \citet{2013A&ARv..21...59I} for a review).
We used the standard energy prescription \citep{Webbink1984} for the CE evolution in our study, that is,
\begin{equation}
E_{\rm b}= \alpha_{\rm CE}(E_{\rm {o,f}}-E_{\rm {o,i}}),
 \end{equation}
where $E_{\rm o,f}$ and $E_{\rm o,i}$ are final and initial orbit energy, respectively.
$E_{\rm b}$ is the binding energy of the CE and is written as
\begin{equation}
E_{\rm b}= \frac{GM_{\rm d}M_{\rm e}}{\lambda R_{\rm d}}
 ,\end{equation}
where $M_{\rm e}$ and $R_{\rm d}$ are the envelope mass and radius of the donor, respectively.
The CE ejection efficiency $\alpha_{\rm CE}$ and structure parameter of the envelope $\lambda$ are highly uncertain.

In our simulations, the $\lambda$ value was computed with the fitting formula of \citet{2014A&A...563A..83C} (see their Appendix A ), 
where $\lambda$ varies with stellar mass, envelope mass, luminosity, and evolutionary stage (i.e., stellar type). The $\alpha_{\rm CE}$ value was assumed to be a constant and was set it to be 0.5, 1, 2, and 3. Being a measure of energy conversion efficiency, values of alpha that are higher than 1 are technically unphysical. However, because other sources of energy in the envelope are not considered, models with values of alpha higher than 1, which reflect the presence of these additional energy sources (see, for example, \citet{Fragos_2019}), tend to perform better in general when tested by observations.

 \begin{figure}
   \centering
   \includegraphics[width=\hsize]{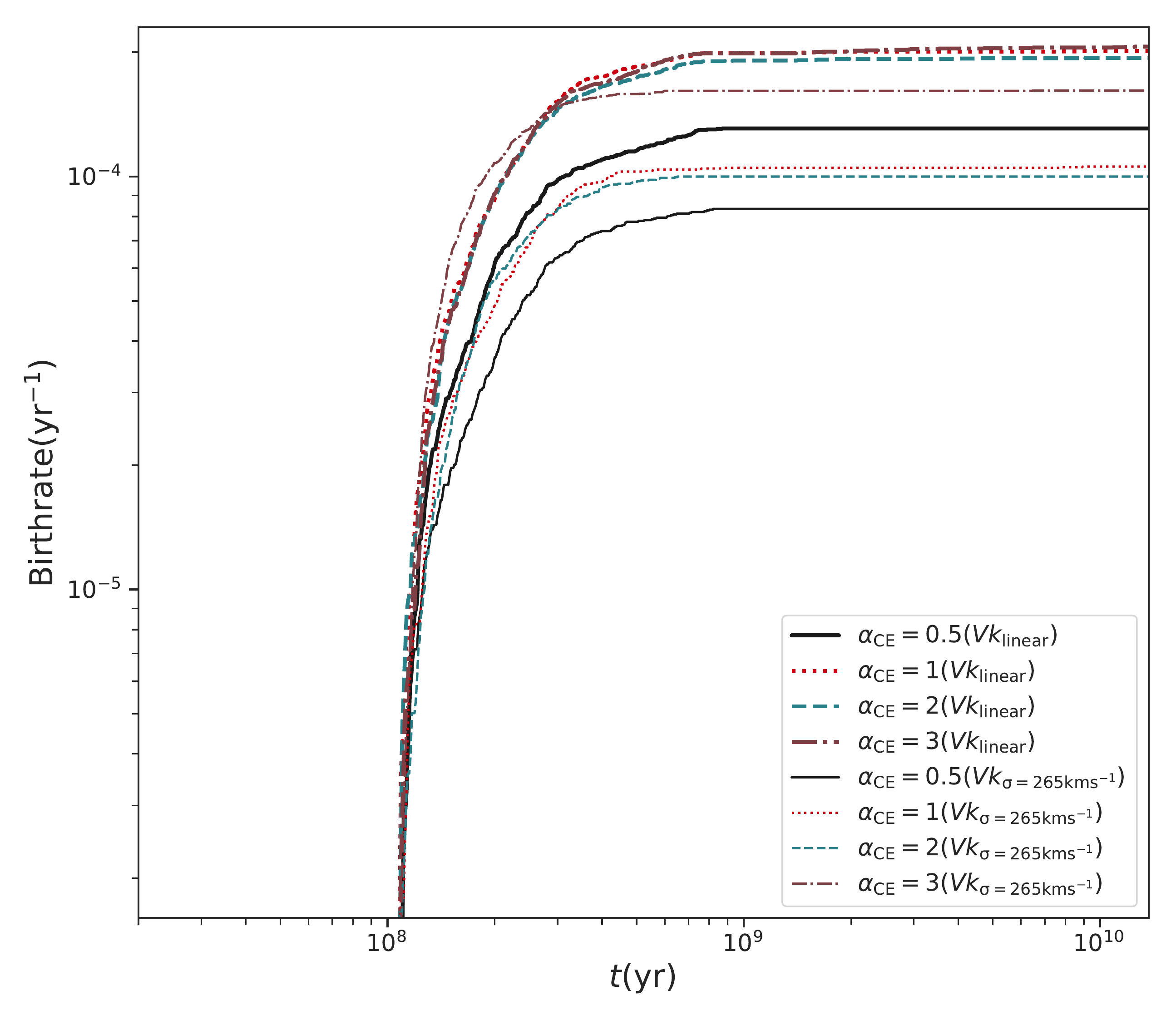}
      \caption{Birthrates of sdB+NS binaries in our simulations
      when a constant Galactic star formation rate of $5M_{\odot}\mathrm{yr^{-1}}$ is assumed.
      Different line styles represent different parameter settings of the simulation.}
         \label{bir}
   \end{figure}

\section{Results}

\subsection{Sample of NS+MS, NS+HG, and NS+GB binaries}
In our simulations,
we first obtained a sample of NS+MS, NS+HG, and NS+GB binaries from BSE, then obtained the sdB+NS binaries
by interpolating in the model grid. The properties of NS+MS, NS+HG, and NS+GB binaries are therefore helpful
for understanding the formation process of sdB+NS binaries.

\subsubsection{Distribution of the companion mass - orbit period}
In Fig.3 we present the distribution of NS+MS, NS+HG, and NS+GB binaries
in the $\log{M_{\rm 2}}$-$\log{P}_{\rm m}$ plane, where $P_{\rm m}$ is the orbital period
when the companion just fills its Roche lobe.
The parameter spaces for producing sdB+NS binaries are overplotted for comparison.

In all the simulation sets,
NS+MS, NS+HG, and NS+GB binaries are from two evolutionary channels.
Binaries with more massive companions ($\log M_{\rm{2}} \geqslant 1.1$) are from the stable RLOF channel,
and the CE ejection channel results in less massive companions ($\log M_{\rm{2}} < 1.1$).
The mass of $M_{\rm 2}$ from the stable RLOF channel is so massive
because they are initially massive to ensure that the mass transfer is stable and that they accrete some material during RLOF. Fig.3 shows that all the NS+MS, NS+HG, and NS+GB binaries from stable RLOF are outside the parameter space and do not contribute to the production of sdB+NS binaries.
Very few NS+MS, NS+HG, and NS+GB binaries (especially in the bottom panels) fall in the parameter space
for producing sdB+NS binaries when $\log M_{\rm2 }  < 0.4$ ($M_{\rm 2}<2.0M_{\odot}$).
This is expected because NSs with such low-mass MS, HG, and GB components are from CE ejection and therefore have relatively short orbital periods, while long orbital periods are required to produce sdB stars from such low-mass progenitors.  
Even if He star+MS binaries with a long orbital period could be produced after CE ejection,
these systems are likely to be destroyed during an SN if the NS natal kick is strong enough,
such as the $Vk_{\rm{\sigma = 265 km s^{-1}}}$ case shown in the bottom panels of Fig.3.
This significantly affects all properties of sdB+NS binaries, as we show below.

 \begin{figure*}
   \centering
   \includegraphics[width=\hsize]{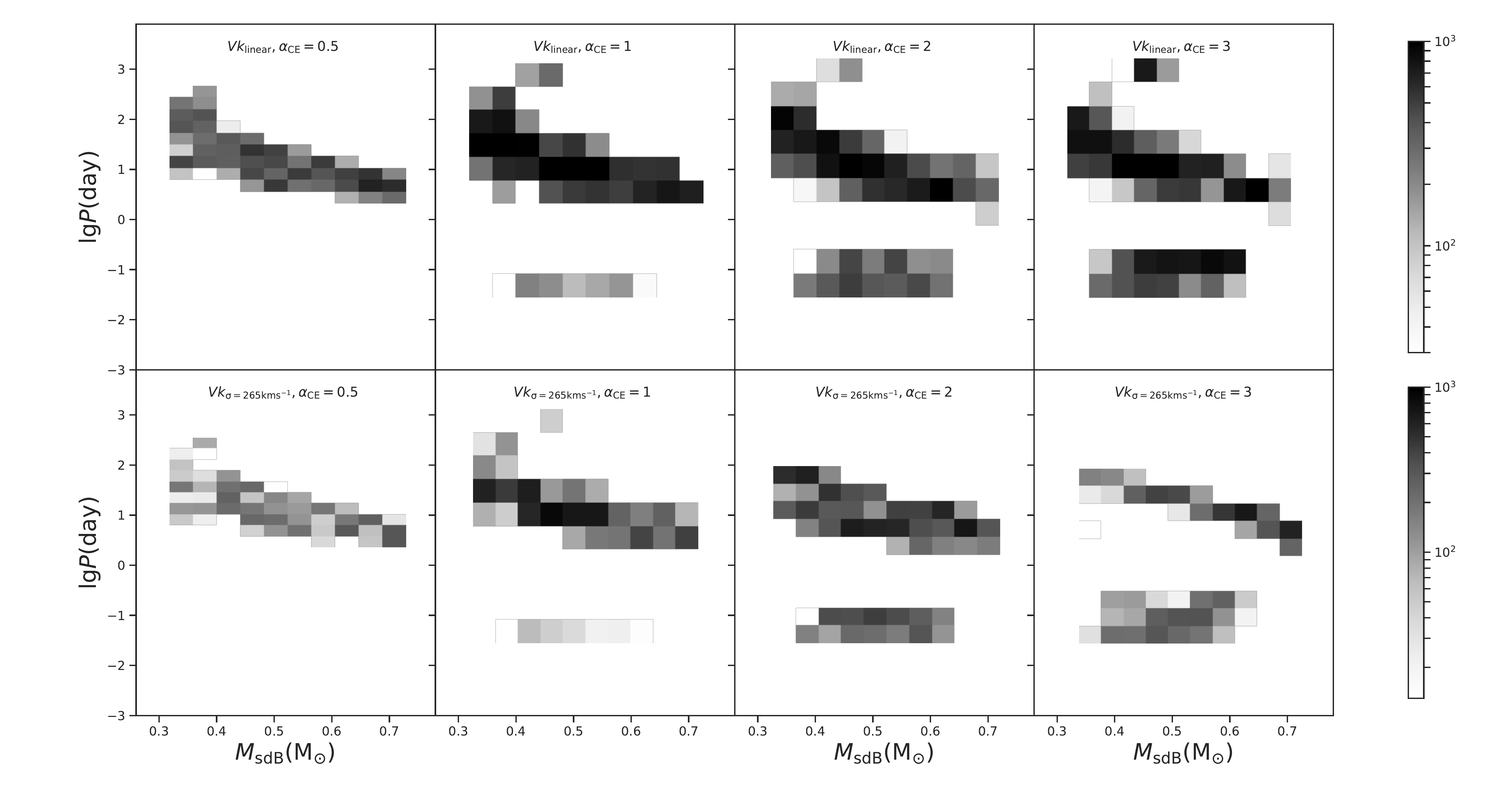}
          \caption{Number distributions in the sdB mass ($M_{\rm sdB}$) - period ($\log P$) diagram for Galactic sdB+NS binaries at the present epoch. The gray scale shows the numbers of each pixel, and the parameter settings for the simulation are indicated in each panel.}
         \label{mnsmsn1}
   \end{figure*}

\subsubsection{Effect of $\alpha_{\rm CE}$}
In binary evolution, the orbital period $P_{\rm m}$ of an NS+MS, NS+HG, and NS+GB binary
produced from CE ejection is affected by two factors:
the value of $\alpha_{\rm CE}$ , and the NS natal kicks.
The higher the $\alpha_{\rm CE}$ value,
the longer the orbital period of He star+MS binaries after the CE ejection.
These He+MS binaries are likely destroyed by SN in the subsequent evolution.
In the case of $Vk_{\rm{\sigma = 265 km s^{-1}}}$, the NS natal kick is independent of the He star mass.
The distribution on the $\log{M_{\rm 2}}$-$\log{P}_{\rm{m}}$ plane then
changes significantly for the different $\alpha_{\mathrm{CE}}$ values,
that is to say, more systems have long $P_{\rm m }$ with increasing $\alpha_{\mathrm{CE}}$.
For the $Vk_{\rm linear}$ description, however, the NS natal kick depends on the He star mass.
The effect from different $\alpha_{\rm{CE}}$ values is offset by NS kick velocities,
and the distribution changes little, as shown in the top panels of Fig.3.

\subsubsection{Effects of NS kicks}
The prescription of NS natal kick has a great effect on the formation of NS+MS, NS+HG, and NS+GB binaries.
The higher kick velocity may result in more binaries being destroyed in SN explosion and in higher eccentricities of orbits after SN explosion if the binaries have not been destroyed.

As shown in the bottom three panels of Fig.3 ($Vk_{\rm{\sigma = 265 km s^{-1}}}$), there is an
 obvious gap for the orbital periods when $\log M_{\rm{2}} \geqslant 1.1$.
This is due to different strengths of the tidal effect on binaries with different orbital periods after SN explosions. 
When binaries have short orbital periods after SN explosion, a strong tidal effect causes the orbit to become circular soon, and the secondary fills its Roche lobe 
as usual. This corresponds to the NS+MS, NS+HG, and NS+GB binaries below the gap in the figure.
However, for binaries with intermediate orbital periods after SN explosions, the tidal effect is not strong enough and the orbits maintain 
high eccentricities for a long time. The binary separation of binaries
at periastron is smaller than the Roche-lobe radius of the secondary
(In the BSE code, the Roche-lobe radius is calculated with the following formula: $\frac{R_{\rm  rl}}{a} = \frac{0.49q^{2/3}}{0.6q^{2/3}+\ln{(1+q^{1/3})}}$,
where $a$ is the semi-major axis instead of the binary separation in an eccentric orbit, and $q$ is the binary mass ratio.).
As the secondary evolves,
its radius increases and engulfs the NS first at periastron before it fills its Roche lobe. In this case, the NS+MS, NS+HG, and NS+GB binaries most likely enter the CE phase and eventually merge. Therefore they do not appear in the figure. Furthermore, for those with long orbital periods after SN explosions, the Roche-lobe radius of the secondary
is smaller than the binary separation at periastron. The MS, HG, and GB companions will fill their Roche lobe first. These binaries correspond to the part above the gap in the future.



The gap disappears in models with the $Vk_{\rm{linear}}$ description (the top three panels in Fig.3). In this case, the kick velocity imparted on the NS is significantly weaker than that from the $Vk_{\rm{\sigma = 265 km s^{-1}}}$ description when $\log M_{\rm{2}} \geqslant 1.1$.
Then the eccentricities accordingly decrease, which avoids the merger of NS+MS, NS+HG, and NS+GB binaries at periastron.

In the $Vk_{\rm{linear}}$ description when $0.7<\log M_{\rm{2}} < 1.1$, another gap occurs.
The NS+MS, NS+HG, and NS+GB binaries in this range stem from the CE ejection channel.
The progenitors of NSs are more massive than those from the stable RLOF with $\log M_{\rm{2}} > 1.1$. The ejected mass during the SNe is therefore relatively high and the NSs receives much higher kick velocities, according to the $Vk_{\rm{linear}}$ description.
For similar reasons as described above,
the NS+MS, NS+HG, and NS+GB binaries with intermediate orbital periods might merge at periastron, leaving the gap in this range.

\subsection{Properties of Galactic sdB+NS binaries}

\subsubsection{Numbers and birth rates}

In order to obtain the birthrates and numbers of sdB+NS binaries of the Galaxy from different simulation sets, we first studied the case of a single starburst, then convolved the binaries with the star formation history of the Galaxy. Here we adopted a constant star formation rate of $5M_{\odot}\mathrm{yr^{-1}}$ \citep[see][]{2004A&A...419.1057W}. When we count the numbers of sdB+NS binaries at the present epoch in the Galaxy, we need to know the lifetime of an sdB+NS binary, $t_{\rm sdB+NS}$. In general, the value of $t_{\rm sdB+NS}$ is taken as the time of the core He burning of the sdB. However, if the merger time from GWR, $t_{\rm merge}$, is shorter than the timescale for core He burning, we adopt $t_{\rm merge}$ as the $t_{\rm sdB+NS}$.

Fig.4 shows the birthrate as a function of age for our Galaxy.
The birthrate sharply increases when the age is just older than $10^8$ yr,
and it is constant after about $(5-6) \times 10^8$ yr. This property is mainly determined by the evolutionary age of the progenitors of sdB stars.

\begin{table}
\caption{\label{t7}Results for various models.}
\centering
\begin{tabular}{ccccc}
 \hline \hline
$\mathrm{Set}$ & $\mathrm{Number}^{(a)}$&$\mathrm{CE+CE}^{(b)}$ & $\mathrm{CE+RLOF}^{(c)}$ & $\mathrm{Birthrate}^{(d)}$\\&&&&$(10^{-4}yr^{-1})$\\
\hline
 1  &12491 & 0\% & 100\%& 1.31              \\
 2  &16120 & 2\% & 98\%& 2.01              \\
 3  &19265  & 23\% & 77\%& 1.94          \\
 4  &21604 & 30\% & 70\%& 2.06          \\
 5  &7510 & 0\% & 100\%& 0.83    \\
 6 &7733  & 3\% & 97\%& 1.06  \\
 7  &8910 & 21\% & 79\% & 1.00 \\
 8  &7748 & 40\% & 60\%&1.62\\
  \hline  \hline
\end{tabular}
\tablefoot{$^{(a)}$ Total estimated number in the Galaxy at the present epoch; $^{(b)}$ percentages of sdB+NS binaries from the $\mathrm{CE+CE}$ channel of the total number; $^{(c)}$ percentages of sdB+NS binaries from the $\mathrm{CE+RLOF}$ channel of the total number; $^{(d)}$ birthrate of sdB+NS binaries at the present epoch.}
\end{table}

The current birthrate of sdB+NS binaries in the Galaxy is about $10^{-4}{\rm yr}^{-1}$,
and the number is $\sim~7000-21000$,
a large portion of which are from the CE+RLOF channel.
The CE+CE channel only contributes little to Galactic sdB+NS binaries
because the lifetimes of sdB+NS binaries in this way are too short, that is, $  {\rm \text{they live some megayears}}$ (see Paper I). With the increase in $\alpha_{\rm{CE}}$ , the orbital periods after CE ejection increase accordingly, and the merger timescale tends to be longer.
Thus the portion of sdB+NS binaries from the CE+CE channel gradually increases.
The birthrate has the highest value of $2.06 \times 10^{-4} {\rm yr}^{-1}$ for set 4,
and it changes little with $\alpha_{\rm{CE}}$ for a given prescription of NS natal kick.

Comparing our results with those of the comprehensive study of \citet{Han2003}, we find that for set 4,
the sdB+NS binaries contribute 0.3-0.5\% to the total number of sdB binaries.
The number differences induced by different prescriptions for NS natal kicks are within a factor of $\sim 2$,
and the linear formula of kick velocity considering the effect of various He star masses before SN results in higher birthrates.

Our study shows that the fraction of sdB+NS binaries from the ECSNe and AIC channel
varies with $\alpha_{\rm{CE}}$. 
This fraction is about 
22-37\%, 35-73\%, 18-40\%, and 6-19\% for the models with $\alpha_{\rm{CE}}=$0.5, 1, 2, and 3, respectively. 
In the model with $\alpha_{\rm{CE}} = 1$, the typical orbital period of NS+MS, NS+HG, and NS+GB binaries from the ECSNe and AIC channel is about 0.5-10\;days. Most of these binaries are located in the parameter space that is optimal for producing sdB+NS binaries. With any change of $\alpha_{\rm{CE}}$ relative to this value, the orbital periods of NS+MS, NS+HG, and NS+GB binaries from the ECSNe and AIC channel changes in a way that increasingly more binaries are located outside of the optimal area in parameter space. The number of sdB+NS binaries emanating from the ECSNe and AIC channel will therefore decrease with any deviation of the $\alpha_{\rm{CE}}$ value from $1$.
The properties of these binaries, such as delay time, mass of the sdBs, and orbital periods, cannot be distinguished from that of the sdB+NS binaries that experienced the CCSNe because the properties of sdB+NS binaries are mainly determined by the progenitors for sdB stars.

\subsubsection{The sdB mass, orbital period, and RV semi-amplitude }
While the $\alpha_{\rm{CE}}$ value has little effect on the birthrate and number of Galactic sdB+NS binaries,
it affects the orbital periods of the sdB+NS binaries.
Fig.5 shows the sdB mass - period ($M_{\rm sdB}$-$\log{P}$) distribution
of Galactic sdB+NS binaries at the current epoch from our simulations.
With the exception of the models with $\alpha_{\rm{CE}}=0.5$, we obtain two groups of sdB+NS binaries for each simulation set:
one (with long $P$, from several to $\sim 1000$ d) from the CE+RLOF channel, and
the other (with short $P$, shorter than $\sim 0.1$ d) from the CE+CE channel.
The distributions are very similar to each other in the upper panels (for $Vk_{\rm{linear}}$),
but display a certain level of divergence among the lower ones (for $Vk_{\rm{\sigma = 265 km s^{-1}}}$). The models with  $\alpha_{\rm{CE}}=0.5$ are an exception because these models only generate sdB+NS binaries that can currently be observed through the CE+RLOF channel.
The greatest difference comes from set 8,
for which the samples from the CE+RLOF channel obviously gather at the long-period end.
This can be understood as follows.
Because of the relatively high $\alpha_{\rm{CE}}$ vaule for set 8 (in comparison to sets 5, 6, and 7),
the NS+MS, NS+HG, and NS+GB binaries produced by the primary binary evolution (the first binary interaction)
have long periods in general, as shown in the bottom right panel of Fig.3.
This leads to relatively long periods for the produced sdB+NS binaries from the CE+RLOF channel in the following evolution, and thus a distribution that is more highly concentrated at the long-period end.

There is a wide range of sdB masses, $\sim 0.32-0.75M_\odot$,
but without a clear mass peak, as shown in Fig.5.
The sdBs from the CE channel have a narrower mass range, but the distribution still does not show a peak.
The sdB+NS binaries from the CE+RLOF channel do not agree with the sdB mass - period relation given by \citet{Chen2013}
because the progenitors for sdB stars are more massive than $2.5M_\odot$ (therefore have non-degenerate He cores) in general, as shown in Fig.3.

Fig.6 shows the RV semi-amplitude $K$ distribution for Galactic sdB+NS binaries,
where the value of $K$ is derived from the binary mass function \citep[e.g.,][]{Geier2011}
by assuming an inclination angle of $90^{\circ}$.
The main peak arises at about $150 \rm {kms^{-1}}$ for the samples from the CE+RLOF channel,
and a small peak lies between $400-600 \rm {kms^{-1}}$ , produced from the CE+CE channel. Again, $\alpha_{\rm{CE}}=0.5$ leads to exceptions, for the same reason as explained above.

 \begin{figure*}
   \centering
   \includegraphics[width=\hsize]{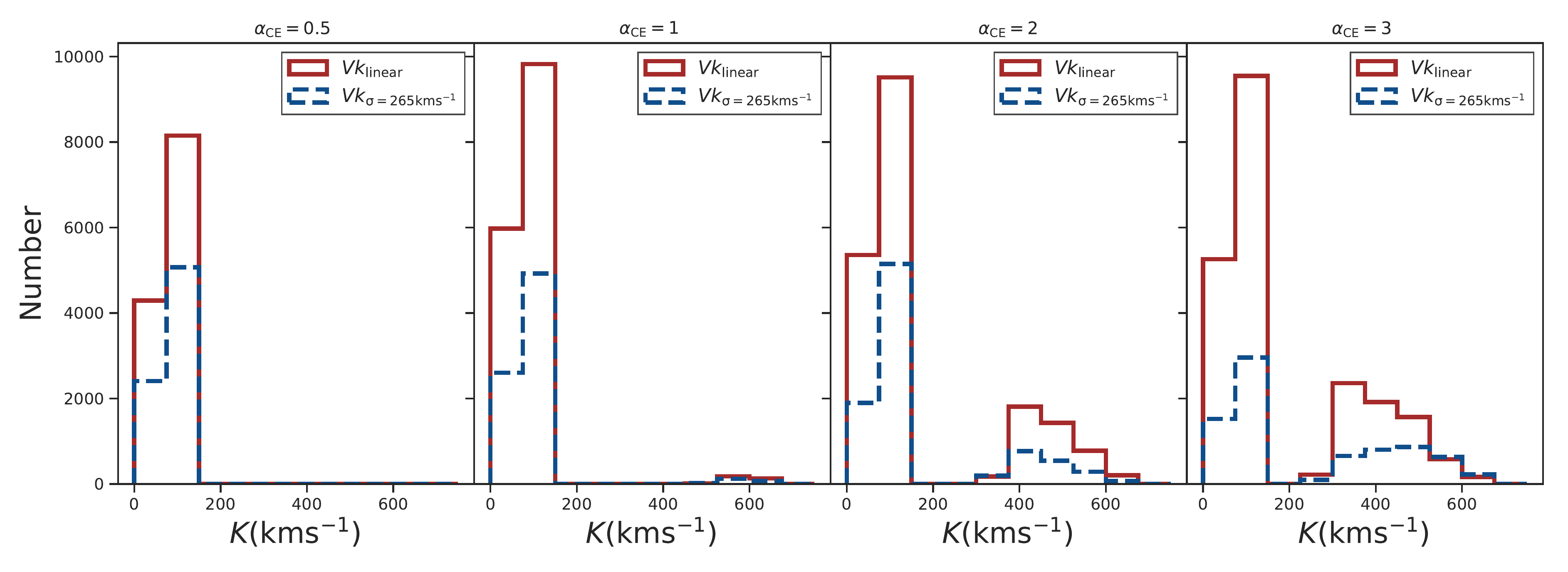}
      \caption{Radial velocity semi-amplitude $K$ distributions of Galactic sdB+NS binaries at the present epoch.
     The value of $\alpha_{\rm CE}$ is described at the top.
     The prescriptions of NS natal kicks, i.e.,
     $Vk_{\rm{linear}}$ (red) and $Vk_{\rm{\sigma = 265 km s^{-1}}}$(blue)
     are indicated in the legend.}
         \label{kkbinn}
   \end{figure*}
 \begin{figure*}
   \centering
   \includegraphics[width=\hsize]{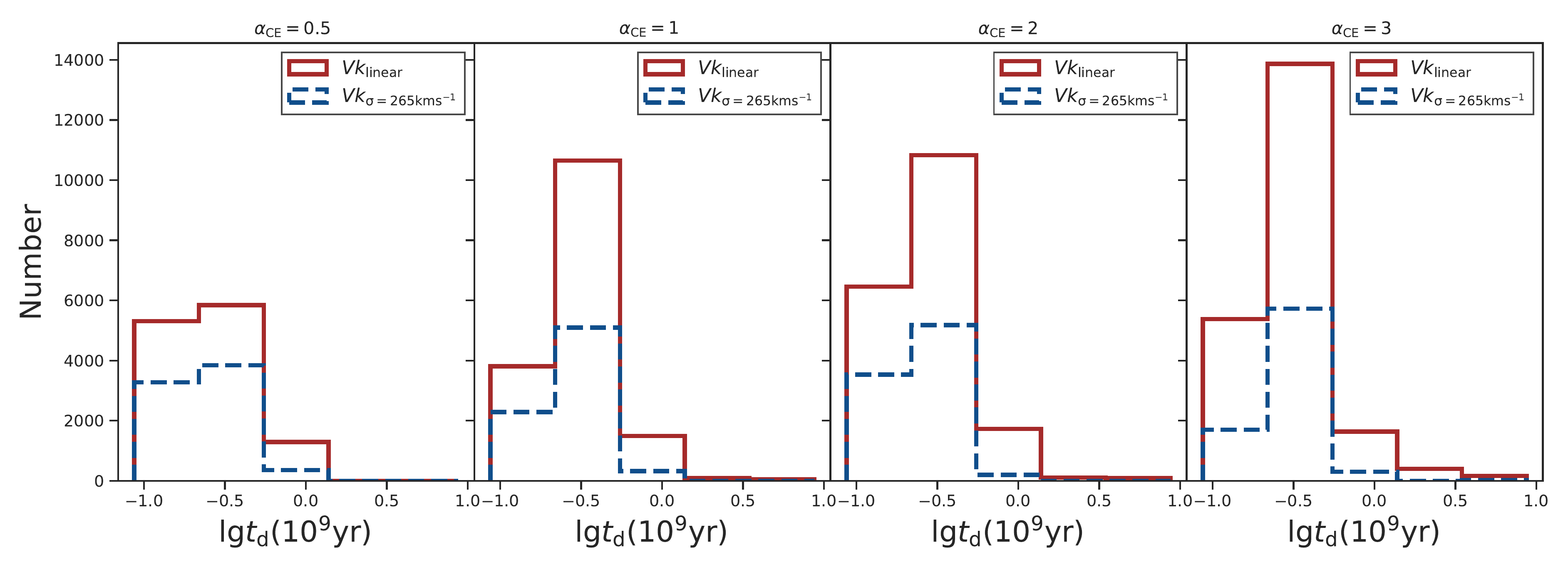}
      \caption{Similar to Fig.6, but for the delay time $t_{\rm{d}}$ (from the primordial zero-age main-sequence binaries to the formation of sdB+NS binaries) distributions of Galactic sdB+NS binaries at the present epoch.}
         \label{ttbinn}
   \end{figure*}
 \begin{figure*}
   \centering
   \includegraphics[width=\hsize]{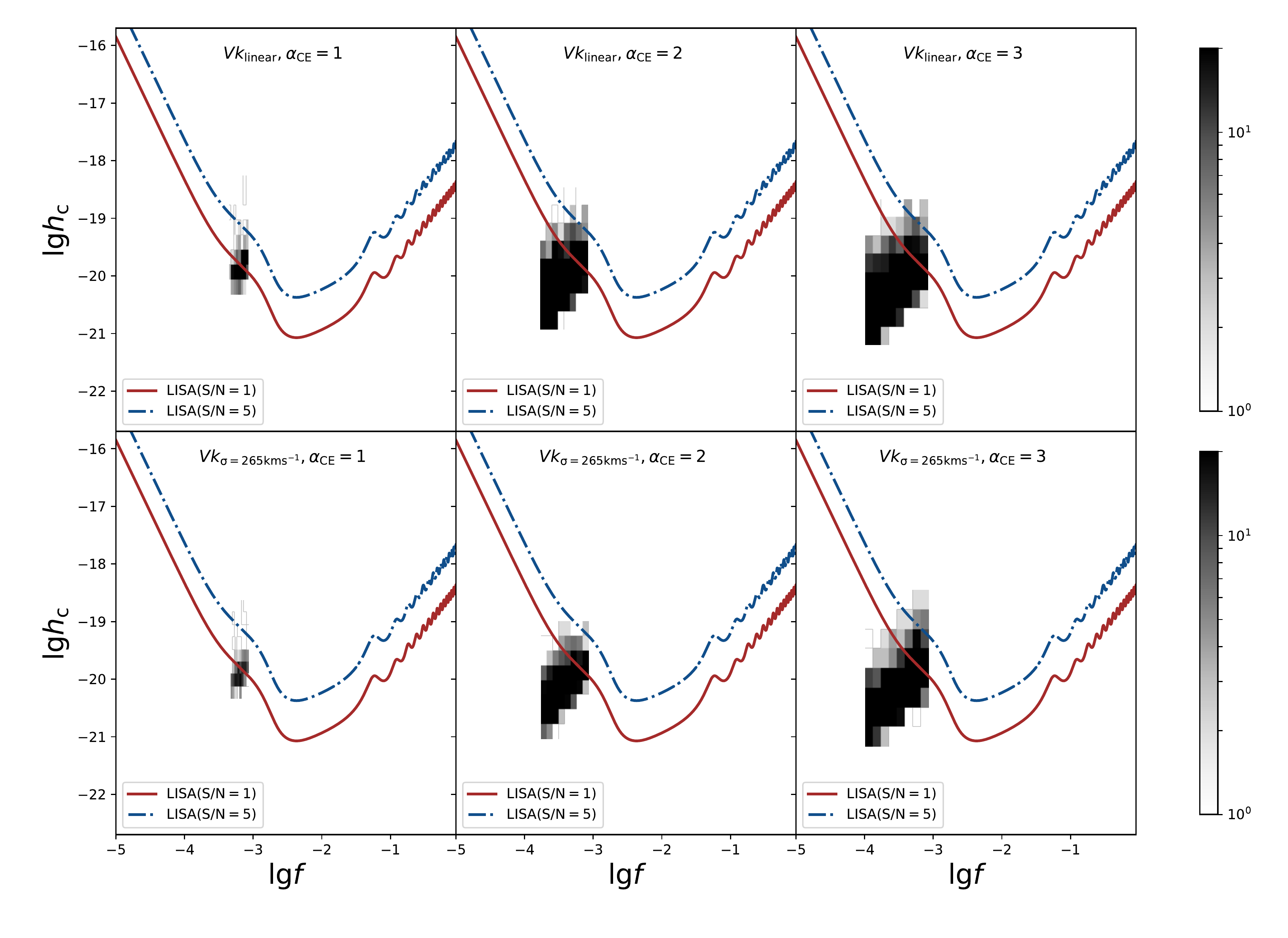}
      \caption{Gravitational wave characteristic strain vs. frequency sdB+NS binaries from the CE+CE ejection channel
      in the Galaxy. The solid and dot-dashed lines are LISA sensitivity curves
      for an S/N of 1 and 5, respectively, with an integration time of 4 yr \citep{2016PhRvD..93b4003K}.}
         \label{bsegww}
   \end{figure*}

\subsubsection{Progenitor ages}
In addition to the characteristics of sdB+NS binaries,
the progenitors also hold some clues for searching them.
Fig.3 has already shown that $M_{\rm 2}$, which is more massive than $2.5M_\odot$, can generally produce sdB+NS binaries from the CE+RLOF channel or the CE+CE channel.
For each sdB+NS binary in the Galaxy, we determined the delay time $t_{\rm d}$ (i.e.,
from the primordial zeor-age main-sequence binaries to the formation of sdB+NS binaries) and show its distribution in Fig.7.
The peak around 0.2 Gyr is obvious, and the majority has delay times shorter than $10^9$ yr.
This indicates that Galactic sdB+NS binaries are born in very young populations, which
supports the idea that sdB+NS binaries are located in the Galactic disk.
The fact that no sdB+NS binaries have been discovered by the MUCHFUSS project is then easily understood
because the candidates for the project are from the SDSS,
and most objects exceeding $\sim$ 3 kpc may be located in the Galactic halo and are old.

\subsection{GWR signals from short-period sdB+NS binaries}
As noted in Paper I,
the sdB + NS binaries produced from the CE+CE channel could be identified as gravitational wave radiation (GWR) sources because their orbital periods are extremely short.
Our simulations show that up to $\sim$6000 sdB+NS binaries can be produced from the CE+CE channel at the present epoch in the Galaxy (see Table 2), depending on the values of $\alpha_{\rm{CE}}$.
The GWR characteristic strains $h_{\rm c}$ versus frequency $f$ for these objects are presented in Fig.8.
The distance for each sample was obtained through a Monte Carlo simulation
with the Galactic potential model of \citet{2016Astraatmadja}, similar to what has been done in \citet{2010Yu}.

Fig.8 shows that part of these sdB+NS binaries ($\sim 100-300$)
radiate GW with strains above the sensitivity curve of the Laser Interferometer Space Antenna (LISA) for  an $S/N=1$. Although the
signals may be covered by confusion-limited background at low frequencies (
The confusion-limited noise background can be produced from double white dwarfs at low frequencies $\sim \lg f < -2.8$ \citep[see][]{2001A&A...375..890N}), systems with a very strong signal might be extracted from the noise \citep{2011PhRvD..84f3009L}.
They may be detectable by LISA in the future.

\section{Conclusions and outlook}

The purpose of this study was to obtain the population properties of Galactic sdB+NS binaries through BPS studies,
such as the number (or birthrate), orbital period, RV semi-amplitude distributions, and the delay times.
Different prescriptions of NS natal kicks and parameters of CE evolution were investigated to
evaluate the uncertainties from SN explosion and from binary evolution.
Our main conclusions are listed below.

(i) There are two mass-transfer processes from a primordial binary that evolves into an sdB+NS binary. The first process (for producing a He star, then an NS) is dynamically unstable.
The second mass-transfer process might be dynamically stable or unstable,
and the low-mass MS, HG, andGB companions ($\le 2.0M_\odot$, with degenerate He cores)
hardly contribute to the formation of sdB+NS binaries because the orbital periods of NS binaries with such low-mass companions are not long enough for the companion to evolve into sdBs.

(ii) The Galactic birthrate of sdB+NS binaries is about  $10^{-4}\mathrm{yr^{-1}}$,
and the highest value is $2.06 \times 10^{-4}\mathrm{yr^{-1}}$ in our simulations (from set 4).
There are  $\sim 7000-21000$ sdB+NS binaries in the Galaxy at the present epoch,
which contributes 0.3-0.5\% of the total sdB binaries for set 4.
The parameters describing the CE evolution have little impact on the results,
while the prescription for NS natal kicks affects the number (and birthrate) by a factor of $\sim 2$.
The prescription that considers the effect of the progenitor mass of NS results in higher birthrates.

(iii) Most of Galactic sdB+NS binaries ($\gtrsim 60\%$) stem from the CE+RLOF channel. The populations from the CE+RLOF channel and from the CE+CE channel show different orbital periods, as expected. This
results in two peaks in the RV semi-amplitude distribution:
$100-150 {\rm {kms^{-1}}}$ for those from CE+RLOF channel, and $400-600 {\rm {kms^{-1}}}$ for those from the CE+CE channel.
However, the delay times for the two populations are similar, that is, they have a peak at about 0.2 Gyr and are no older than 1.0Gyr. This indicates that Galactic sdB+NS binaries are born in very young populations.
We therefore suggest that most sdB+NS binaries may be located in the Galactic disk and have a small RV semi-amplitude.
The observation of the MUCHFUSS project can be easily understood. It might still be difficult to discover the binaries in the Galactic disk because the $K$ values are low and the extinction is high.

(iv) We find an upper limit of $\sim 6000$ for the number of sdB+NS binaries that evolved through the CE+CE channel in the Galaxy.
They are potentially strong GWR sources. About $\sim 100-300$ might be detected by LISA 4 yr observations with an $S/N=1$.

The results for sdB+NS binaries have several important implications for future studies.
For instance, during the formation of the sdB+NS binaries,
the NS can be spun up to be a millisecond pulsar from the stable RLOF. Moreover, carbon-oxygen ultra-compact X-ray binary might be formed from sdB+NS binaries with short orbital periods. Further studies regarding the subsequent evolution phase of sdB+NS binaries are warranted.

\section{Acknowledgements}

This work is partly supported by the Natural Science Foundation of China (Nos.11733008, 11521303,11703081), the National
Ten-thousand talents program,
CAS light of West China Program and the Youth Innovation Promotion Association of the CAS (grant 2018076).


\bibliographystyle{aa}
\bibliography{My_Collection}
\end{document}